\documentclass{icrc29}
\usepackage{graphicx,amssymb,amsmath,times}
\usepackage[tight]{subfigure}
\begin{document}
\title{Development of acoustic devices for ultra-high energy neutrino
  detectors}
\author[T. Karg et al.]{T. Karg, G. Anton, K. Graf, J. H\"o\ss l,
  A. Kappes, U. Katz, R. Lahmann,
  \newauthor
  C. Naumann, K. Salomon, S. Schwemmer
  \newauthor
  On behalf of the ANTARES collaboration \\
  Physikalisches Institut, Friedrich-Alexander-Universit\"at
  Erlangen-N\"urnberg, \\
  Erwin-Rommel-Stra\ss e 1, 91058 Erlangen, Germany}
\presenter{Presenter: T. Karg (Timo.Karg@physik.uni-erlangen.de),
  ger-karg-T-abs1-og27-poster}

\maketitle

\begin{abstract}
  Acoustic neutrino detection is a promising approach to instrument
  the large detector volumes needed for the detection of the small
  neutrino fluxes expected at ultra-high energies ($E \gtrsim 1 \,
  \mathrm{EeV}$). We report on several studies investigating the
  feasibility of such an acoustic detector. High-precision lab
  measurements using laser and proton beams aiming at the verification
  of the thermo-acoustic model have been performed. Different types of
  acoustic sensors have been developed and characterised. An
  autonomous acoustic system, attached to the ANTARES prototype string
  ``Line0'', has been deployed and operated successfully at $2400 \,
  \mathrm{m}$ depth, allowing for in-situ studies of the acoustic
  background in the Mediterranean Sea.
\end{abstract}

\section{Introduction}

Neutrino fluxes at ultra-high energies (e.g. GZK neutrinos) are
predicted to be very small. Thus large target masses are required to
measure at least a few events during the lifetime of a typical
experiment. The photo-sensor distance in water \v{C}erenkov neutrino
telescopes is limited by the attenuation length of the \v{C}erenkov
light in water or ice ($50$ -- $70 \, \mathrm{m}$), constraining the
affordable size of such detectors.

Another approach to neutrino detection, allowing for detectors
instrumented more sparsely, is acoustic detection which was first
described in \cite{Askariyan}. The particle cascade produced by the
neutrino interaction heats up the medium locally and leads to fast
expansion and a bipolar pressure pulse with a typical frequency of $20
\, \mathrm{kHz}$, which propagates perpendicular to the cascade axis.
This unique disc-shaped event signature allows for good direction
reconstruction and background suppression. The sensor density in an
acoustic detector is determined by the sonic attenuation length in
water, which is about ten times larger than the optical attenuation
length.

It is planned to deploy several acoustic sensors as part of the
ANTARES neutrino telescope \cite{Antares} to study the technical
feasibility and environmental and background conditions of an acoustic
neutrino detector. ANTARES is being installed in the Mediterranean
Sea, $40 \, \mathrm{km}$ off the coast of Toulon.

\section{Tests of the thermo-acoustic model}

The thermo-acoustic model describes the hydrodynamic sound generation
of a particle cascade in water. The resulting pressure field is
determined by the spatial and temporal distribution of the deposited
energy, by the sound velocity, and by the heat capacity and the
expansion coefficient, the latter two depending on temperature. Thus
it is possible to test the model in the laboratory by using other
means of energy deposition.  This has been done with accelerator
proton beams in the past \cite{Sulak, Hunter, Albul}.

We have performed experiments using a pulsed $1064 \, \mathrm{nm}$
Nd:YAG laser, and the $177 \, \mathrm{MeV}$ proton beam of the Gustaf
Werner Cyclotron at the Theodor Svedberg Laboratory in Uppsala,
Sweden.  The beams were dumped into a $150 \times 60 \times 60 \,
\mathrm{cm}^3$ water tank, where the acoustic field could be measured
with several position-adjustable hydrophones, both commercial and
custom-designed.  The temperature of the water could be varied between
$1$ -- $50^\circ \mathrm{C}$ with a precision of $0.1^\circ
\mathrm{C}$ by cooling and subsequent controlled, homogeneous
reheating of the whole water volume.

The spill energy of the proton beam was varied from $10$ to $400 \,
\mathrm{PeV}$. The beam diameter was $\approx 1 \, \mathrm{cm}$ and
the spill time $30 \, \mathrm{\mu s}$. The laser pulse energy can be
adjusted between $0.1$ and $10 \, \mathrm{EeV}$ at a beam diameter of
a few millimetres. The pulse length is fixed at $10 \, \mathrm{ns}$.

\begin{figure}[h]
  \begin{center}
    \subfigure[Laser.]{
      \includegraphics*[height=4.5cm,angle=0,clip]{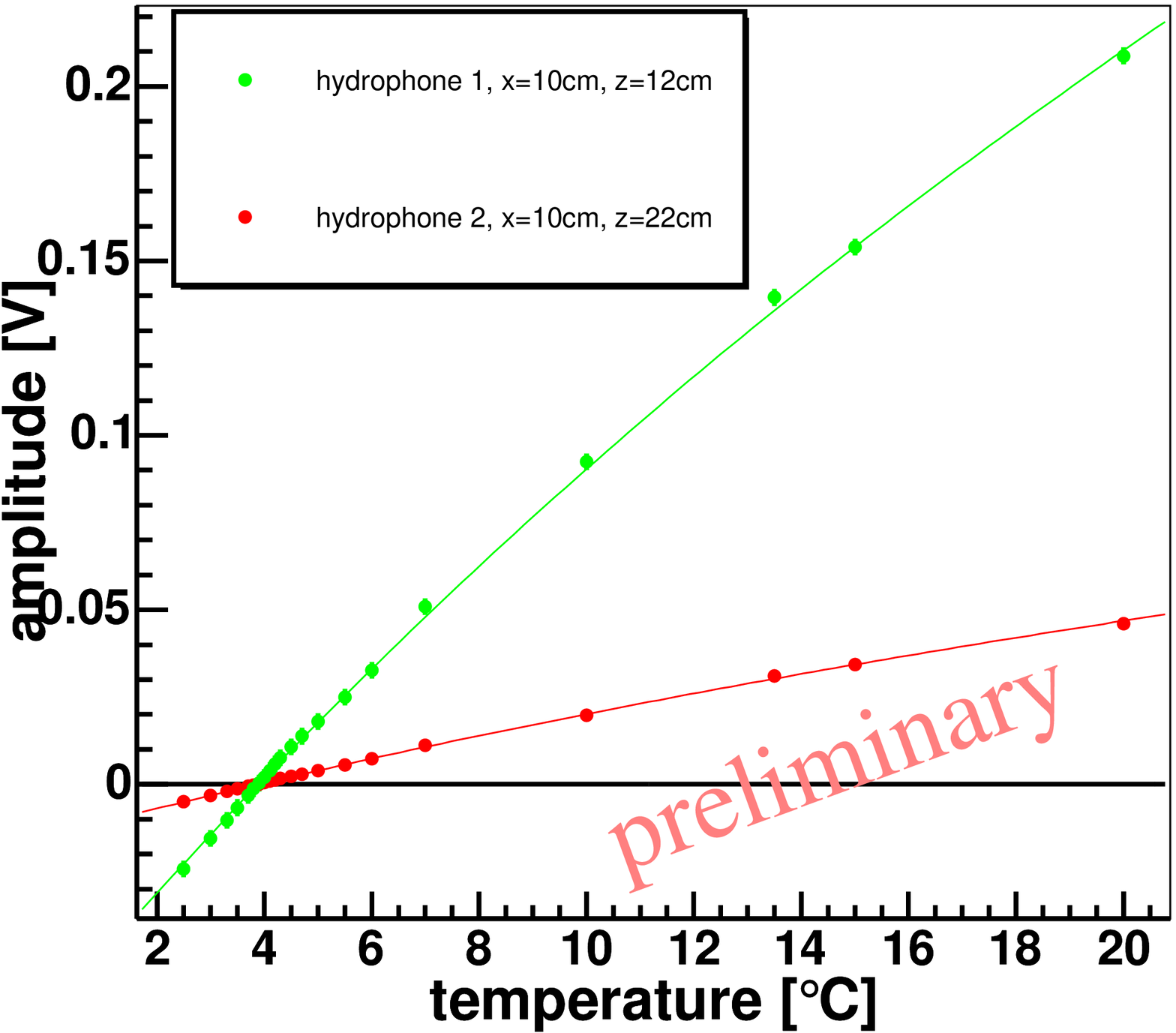}
      \label{fig1a}
    }
    \subfigure[Proton beam.]{
      \includegraphics*[height=4.5cm,angle=0,clip]{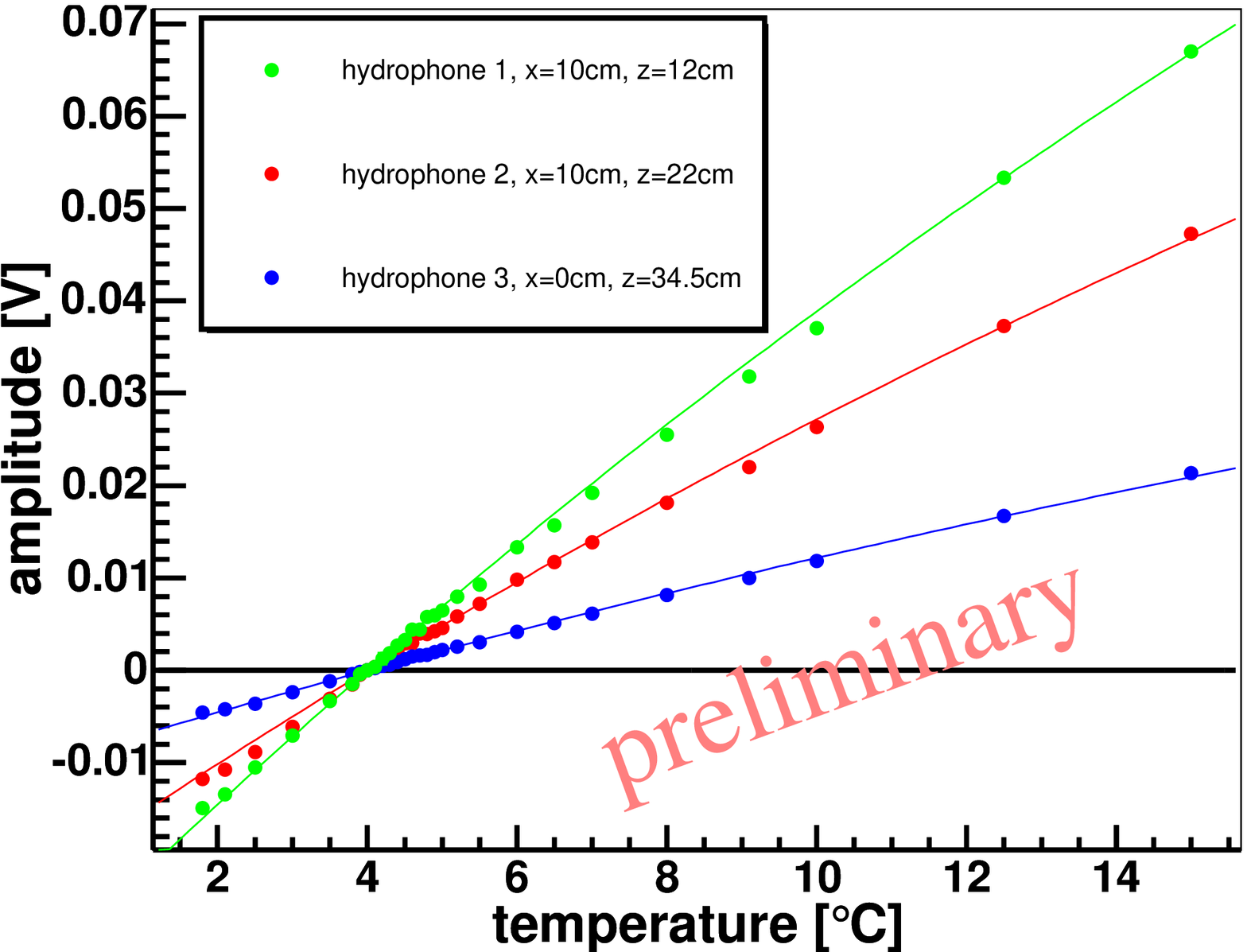}
      \label{fig1b}
    }
    \caption{\label{fig1} Dependence of the peak-to-peak amplitude of
      the bipolar signals on temperature for \subref{fig1a} the laser
      and \subref{fig1b} the proton beam. Inverted signals below
      $4^\circ \mathrm{C}$ have a negative amplitude. The size of the
      error bars is within the size of the data points. Hydrophones at
      different positions relative to the beam are shown ($x$ axis
      perpendicular to beam direction; $z$ axis in beam direction).}
  \end{center}
\end{figure}

The measured bipolar signals are in excellent agreement with
simulations made under the assumption of a thermo-acoustic signal
generation mechanism. Figure~\ref{fig1} shows the temperature
dependence of the peak-to-peak amplitude.  As expected from the
thermo-acoustic model, the laser beam signal shown in
figure~\ref{fig1a} changes its polarity around $4 ^\circ \mathrm{C}$.
The model expectation for the signal amplitude, which is proportional
to the volume expansion coefficient ($\alpha$ vanishes at $4^\circ
\mathrm{C}$), is fitted to the experimental data, using an overall
scaling factor, and a constant temperature shift as free parameters.
This fit yields a zero-crossing of the amplitude at $(3.90 \pm
0.02)^\circ \mathrm{C}$ which is compatible with the expectation of
$(4.0 \pm 0.1)^\circ \mathrm{C}$, where the error is dominated by the
systematic uncertainty in the temperature setting.  Analysing the
proton data in the same way yields a shape slightly deviating from the
model expectation, and a zero-crossing significantly different from
$4^\circ \mathrm{C}$. In view of the results from the laser beam
measurements, we subtract the residual signal at $4.0^\circ
\mathrm{C}$, which has an amplitude of approx.  $1 \, \mathrm{mV}$,
from all signals, assuming a non-temperature dependent effect on top
of the thermo-acoustic signal. The resulting amplitudes shown in
figure~\ref{fig1b} are then in good agreement with the model
prediction. Also signal dependencies on beam energy, beam width and
sensor distance from the beam were investigated, and show very good
agreement with the simulation and expectation from the thermo-acoustic
model.  Thus the thermo-acoustic model could be verified to high
accuracy.

\section{Development and characterisation of acoustic sensors}

It is essential for acoustic neutrino detection to understand the
sensitivity and frequency response of the sensors used.  We have built
two different types of acoustic sensors: piezo ceramics elements, that
are coated with polyurethane, or mounted into pressure tight vessels.
Their electronic and acoustic properties have been investigated. The
second option is particularly interesting when using glass spheres or
titanium vessels that are housing the ANTARES components, since their
water tightness is guaranteed, and they are easy to integrate into the
existing ANTARES detector setup.

\begin{figure}[h]
  \begin{center}
    \subfigure[Impedance.]{
      \includegraphics*[height=4.5cm,angle=0,clip]{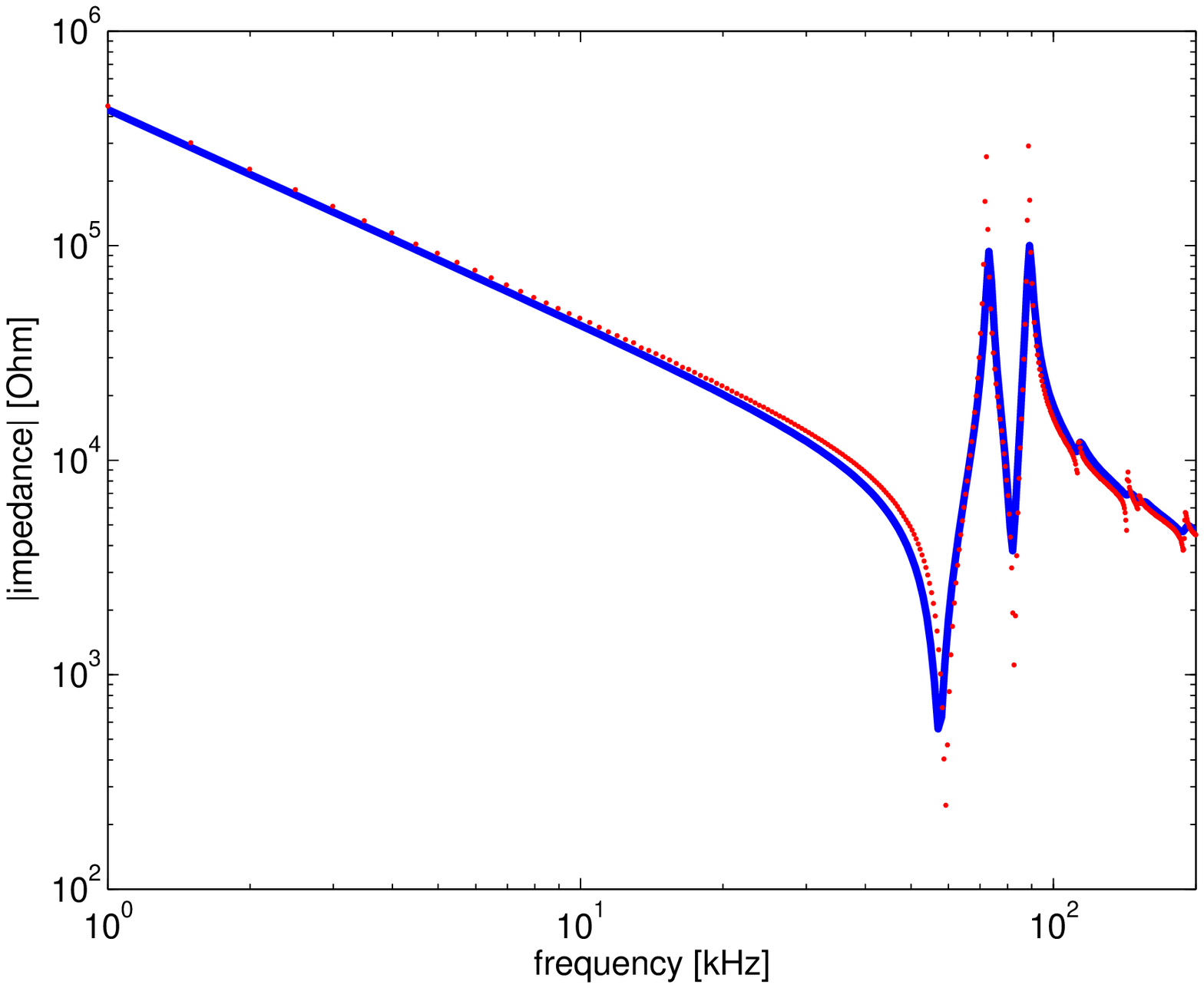}
      \label{fig2a}
    }
    \subfigure[Displacement amplitude.]{
      \includegraphics*[height=4.5cm,angle=0,clip]{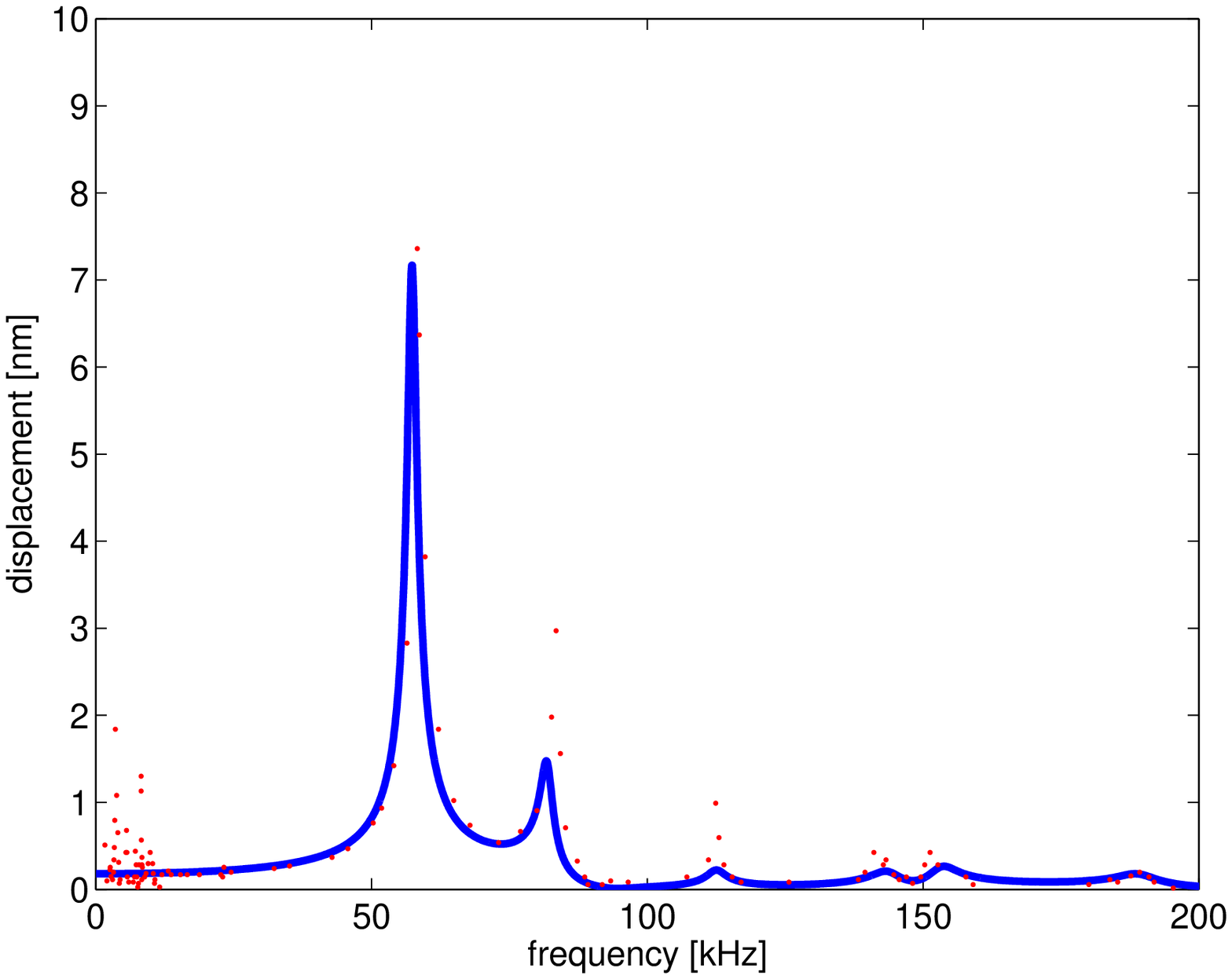}
      \label{fig2b}
    }
    \caption{\label{fig2} Measured frequency dependence \subref{fig2a}
      of the impedance of a piezo ceramic element (red dots) compared
      to predictions from the mechano-electrical model (blue line),
      and \subref{fig2b} of the displacement amplitude of the piezo
      ceramics surface. In both plots the resonances from the piezo
      element are well visible.}
  \end{center}
\end{figure}

Using finite element methods, the frequency-dependent impedance of a
piezo ceramic element can be calculated from the known quantities
permittivity, elasticity modulus, and piezoelectric modulus.  We
measure the impedance by applying white noise to the piezo ceramics
and analysing the frequency spectrum of the voltage measured at a
capacitor in serial connexion. Figure~\ref{fig2a} shows that with this
method, the electrical properties of a piezo can be predicted very
well, allowing for the design of piezo element geometries that are
well matched to the needs of acoustic particle detection.

Further, interferometric measurements of the dependence of the
displacement of the piezo ceramics surface on the applied voltage have
been carried out, that allow for direct calibration of the piezo
ceramics sensitivity to external pressure signals. A laser beam is
coupled into an optical fibre and exits through a fibre end, which is
prepared carefully to behave like a mirror. The beam is reflected
multiple times between this end and a thin, gold coated glass plate a
few micrometres away, that is mounted onto the piezo ceramics surface.
Both surfaces act as a Fabry-Perot interferometer. The light, that, at
each reflection on the fibre end, is coupled back into the fibre, is
measured behind a beam splitter, allowing for the determination of
distance variations between the fibre end and the piezo ceramics
surface. In figure~\ref{fig2b} excellent agreement between the
measurement and finite element calculations based on material
properties can be seen.

The properties of ``naked'' piezo ceramics can be well described with
the methods presented, thus piezo ceramics with high sensitivity for
applications like acoustic particle detections can be designed. The
properties of coated piezo ceramics or piezo ceramics coupled to
pressure tight vessels are under study.

\section{AMADEUS -- \underline{A}utonomous \underline{M}odule for
  \underline{A}coustic \underline{DE}tection \underline{U}nder the
  \underline {S}ea}

\begin{figure}[h]
  \begin{center}
    \subfigure[View into the AMADEUS cylinder from the top.]{
      \includegraphics*[height=4.5cm,angle=0,clip]{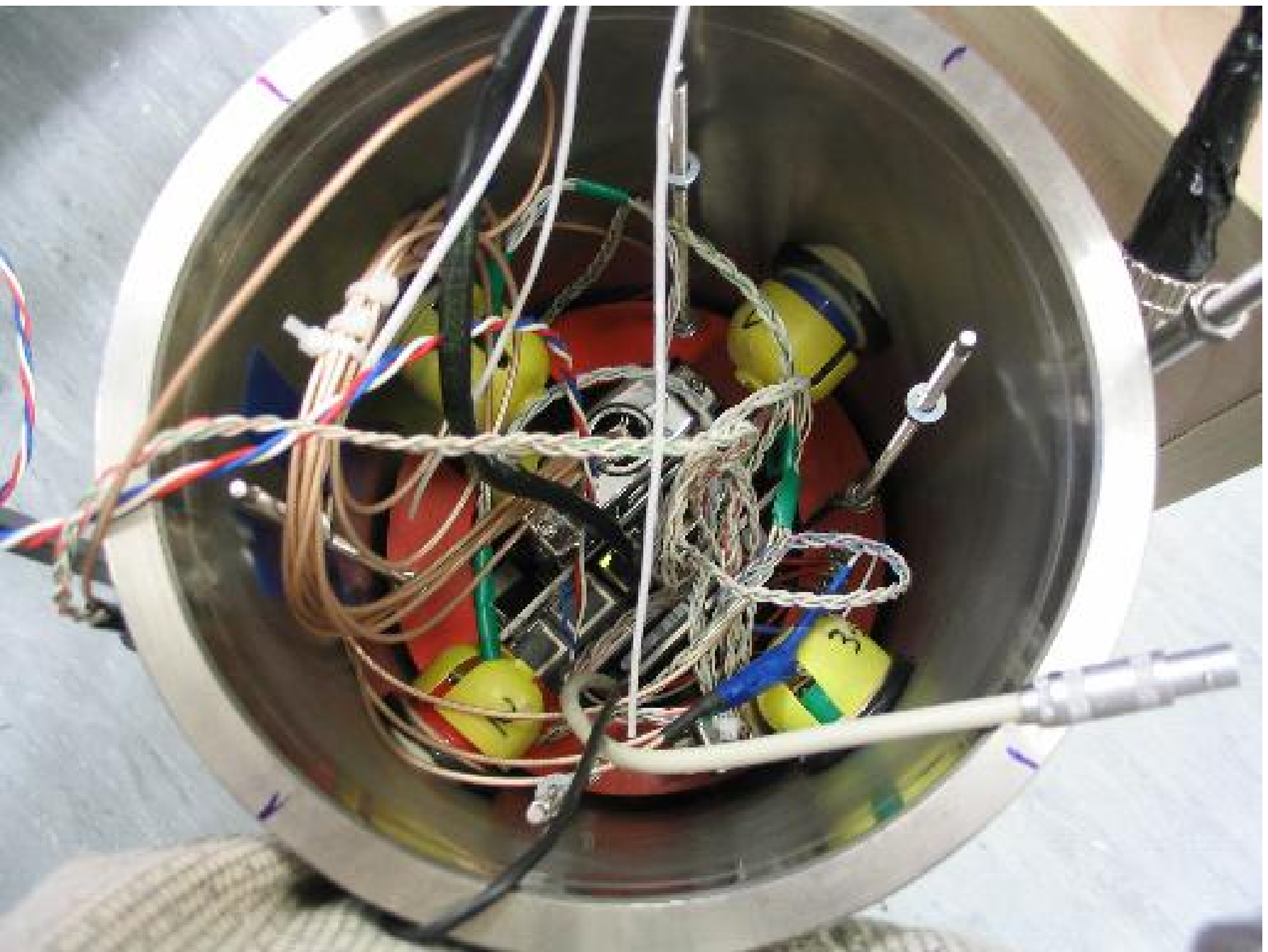}
      \label{fig3a}
    }
    \subfigure[Noise spectrum.]{
      \includegraphics*[height=4.5cm,angle=0,clip]{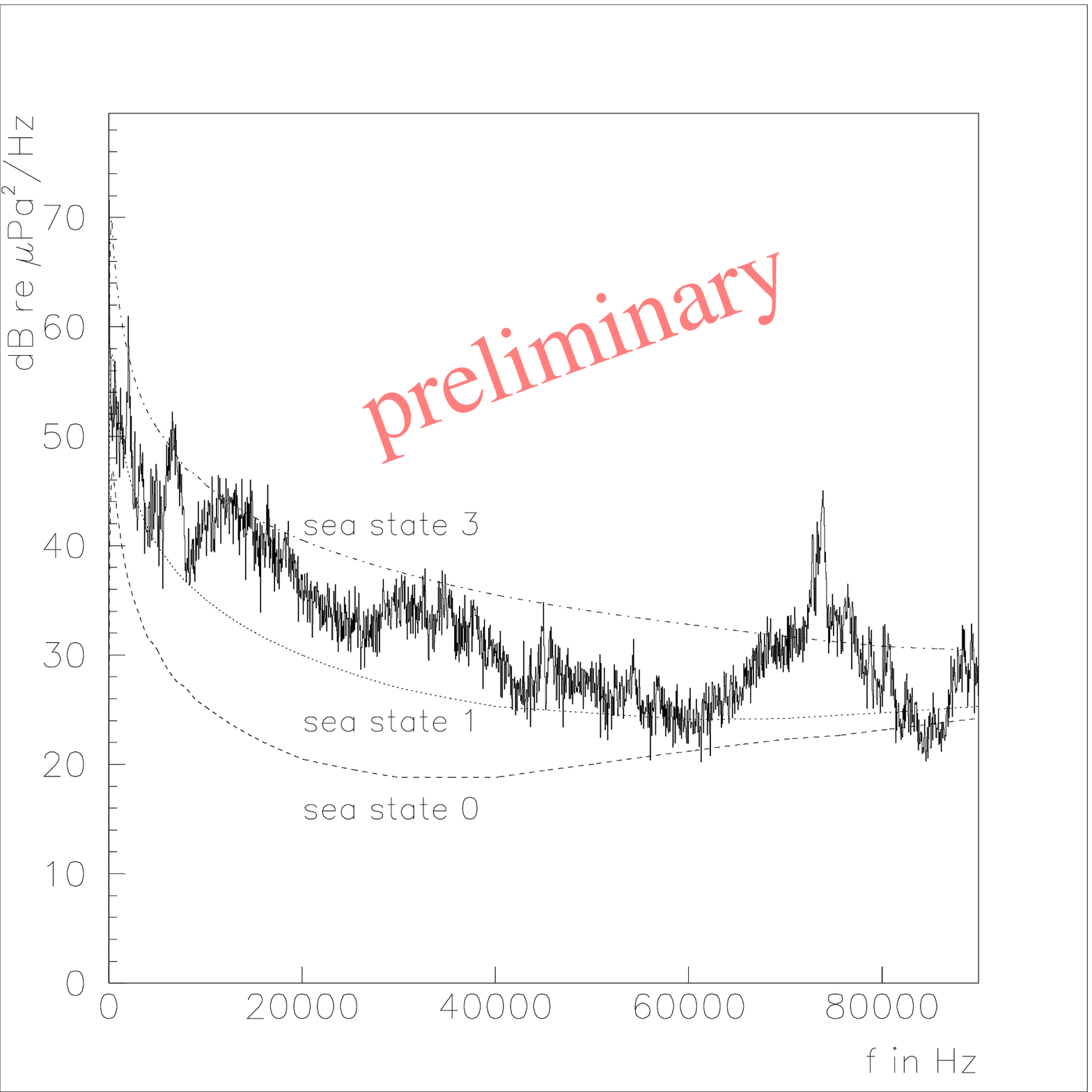}
      \label{fig3b}
    }
    \caption{\label{fig3} In \subref{fig3a} the four yellow sensors
      glued to the walls of the cylinder can be seen; the fifth sensor
      is mounted onto the top cover. The noise power density in
      \subref{fig3b} fits nicely between the predictions for sea
      states one and three (taken from \cite{Urick}).}
  \end{center}
\end{figure}

In spring 2005 an autonomous acoustic data acquisition system
(AMADEUS) was deployed and operated successfully at a depth of $2400
\, \mathrm{m}$ together with the ANTARES prototype test string
``Line0'' in order to study the acoustic background at the ANTARES
site. It consisted of five piezo ceramic sensors glued to the inside
of a pressure tight titanium cylinder normally used for the ANTARES
electronics (cf. figure~\ref{fig3a}). The sensors where read out using
a $16$-bit ADC board with an integral sampling rate of $500 \,
\mathrm{kHz}$. In order to suppress noise from the hard disk, data was
stored on a flash card, and written to disk when no data was recorded.
In different operation modes it was possible to read out a single
sensor with a high data-rate, or multiple sensors in coincidence. A
total of 12 hours of acoustic data were taken.

The analysis of a first part of the data, which was recorded during a
first descent of the line, a short period on the sea-floor, and the
recovery operation, shows the usability of the concept to couple piezo
ceramic sensors to the inside of a pressure tight support structure in
order to detect acoustic signals in the deep sea. Figure~\ref{fig3b}
shows a preliminary noise spectrum measured at the bottom of the
Mediterranean sea, which is in good agreement with expectations.

\vspace*{-3mm}
\section{Conclusions}

We have demonstrated, that the thermo-acoustic model describing the
sound generation of high-energy particle cascades is understood up to
high precision, which is imperative in order to build a neutrino
telescope based on acoustic detection. As a first step towards such a
detector, different types of acoustic sensors were investigated and
their properties were predicted from basic principles, allowing for
the development of tailor-made sensors. Acoustic background data was
collected in-situ in the Mediterranean sea at a depth of $2400 \,
\mathrm{m}$.

\smallskip This work was supported by the German BMBF Grant No. 05
CN2WE1/2.

\end{document}